\documentstyle{article}

\begin{document}

\title{$Ab-initio$ No-Core Shell Model With Many-Body Forces}

\author{M.S. Fayache$^{1}$, J.P. Vary$^2$, B. R. Barrett$^3$, P. Navr\'{a}til$^4$
and  S. Aroua$^{1}$\\
(1) D\'{e}partement de Physique, Facult\'{e} des Sciences de Tunis, \\
Tunis 1060, Tunisia\\
\noindent (2) Department of Physics and Astronomy, Iowa State University\\
Ames, Iowa 50011\\
\noindent (3) Department of Physics, University of Arizona\\
Tucson, Arizona 85721\\
\noindent (4) Lawrence Livermore National Laboratory, L-414,\\
P. O. Box 808, Livermore, CA 94551  \\
}
\date{\today}
\maketitle

\begin{abstract}
Working in the framework of the $ab-initio$ no-core shell model, we 
derive two-body effective 
interactions microscopically for specific harmonic-oscillator 
basis spaces from the
realistic  Argonne $V8'$ 
nucleon-nucleon 
potential. However, our two-body effective interaction is only the 
dominant part of the 
$A-$body ($A \ge  3$) 
effective interaction, which also contains three-, four- and, in general, 
up to $A-$body components. 
We model these $multi-body$ components as a simple two-parameter, 
basis-space-dependent interaction. 
We then investigate the coupling needed to provide accurate 
descriptions of selected nuclear observables. We successfully model 
the multi-body interaction as a zero-range two-body operator times
the number operator
\^ N$_{sps}$, that measures the harmonic-oscillator quanta of the spectator
nucleons. 

\end{abstract}

\section{Introduction}

Despite the tremendous progress realized in the study of the bare 
nucleon-nucleon ($NN$) interaction as well as equally impressive advances in 
the theory of the effective $NN$ interaction in the shell model, it has not been 
possible to systematically account for the properties of finite nuclei solely on the 
basis of two-body interactions. Studies of the roles played by
many-body forces  ($e.g.$ 3-body and 4-body) suggest that they
provide small yet important contributions
\cite{TM, CH}. However, more careful analysis of nuclear many-body systems 
is needed to fully identify these roles \cite{AU}.  
Adding to this complexity, the
$ab-initio$ No Core Shell Model ($NCSM$) \cite{PRC95, Petr96, PRLC00} generates
effective many-body forces which have been shown to decrease in strength with
increasing basis space size. Overall, the nature of the "real" and the "effective" 
many-body forces remains quite complicated and elusive. 

In this work, we present a phenomenological model of three-body 
and higher multi-body 
interactions as a simple, two-parameter and basis-space-dependent interaction. 
This multi-body interaction is added to our theoretically-derived two-body 
effective interaction, and its two parameters are adjusted in each basis space 
so as to reproduce results in agreement with experimental data. Thus, 
ours is a $residual$ multi-body interaction obtained in a 
phenomenological way. Our chosen functional form is a delta function reminiscent of the 
recent attention of Effective Field Theory ($EFT$) to expansions of nuclear interactions 
beginning with zero-range terms. In $EFT$ there is an appropriate cutoff in momentum 
transfers, corresponding to the poorly-known short-range part of the nuclear potential 
\cite{EFT}. In a similar spirit, we model the poorly known many-body forces as
harmonic-oscillator-cutoff-dependent zero-range interactions \cite{lepage}. 

More specifically, we model the residual multi-body interaction 
by a one-parameter two-body delta
interaction and a one-parameter multi-body interaction. The multi-body
interaction consists of a two-body delta times a number operator, \^ N$_{sps}$, 
that measures
the excitations in oscillator quanta of the spectators. We 
expect and find 
that the contribution of the two-body part of the interaction 
weakens as the basis space increases and as our 
theoretically-derived effective two-body 
interaction approaches the exact two-body interaction. On the other hand, 
we find as expected that the multi-body part of our residual interaction remains relatively 
strong, which provides an indication of the role played by $n-$body forces ($n > 2$) in nuclei. 

\section{The Shell-Model Hamiltonian and the Effective Interaction}

The $ab-initio$ approach in shell-model studies of the nuclear 
many-body problem \cite{PRC95, Petr96, PRLC00} begins with the 
one- and two-body Hamiltonian for the $A$-nucleon system, i.e., 

\begin{equation}
H= \sum_{i=1}^{A} \frac{\vec{p_i}^2}{2m} + \sum_{i<j}^{A} V_{ij}
\end{equation}

\noindent with $m$ the nucleon mass and $V_{ij}$ the $NN$ 
interaction.  We then add the 
center-of-mass harmonic-oscillator ($HO$) potential 
$\frac{1}{2}m \Omega^2 \vec{R}^2$, 
with \newline $\vec{R}=(1/A)\sum_{i=1}^{A} \vec{r_i}$. 
The total center-of-mass harmonic-oscillator hamiltonian, 
$H^{\Omega}_{c.m.}=\vec{P}^2_{c.m.}/2Am + 
\frac{1}{2}Am \Omega^2 \vec{R}^2$, 
$\vec{P}_{c.m.}=\sum_{i=1}^{A} \vec{p_i}$,
is later subtracted to leave a purely intrinsic Hamiltonian. 
At this stage we have a convenient mean field acting on each 
nucleon that helps our overall convergence when working in a 
$HO$ set of basis states. We emphasize that we adopt proper measures 
to ensure that the $intrinsic$ properties of the many-body 
system are not affected by the center-of-mass term and we 
discuss our specific procedure for this in more detail below. The 
modified Hamiltonian, thus, acquires a  dependence on the 
$HO$ frequency $\Omega$, and can then be written as

\begin{equation}
H^{\Omega}= \sum_{i=1}^{A} \left [ \frac{\vec{p_i}^2}{2m} + \frac{1}{2}m 
\Omega^2 \vec{r_i}^2 \right ] 
+ \sum_{i<j}^{A} \left [ V_{ij} - \frac{m \Omega^2}{2A}(\vec{r_i} - 
\vec{r_j})^2 \right ]
\end{equation}

\noindent Actual shell-model calculations can only be carried out in a finite  
basis space defined by a 
projection operator $P$, with the complementary space to the finite basis
space (i.e. the excluded space) 
defined by the projection operator $Q=1-P$. Furthermore, due to its strong 
short-range part, the realistic 
nuclear interaction in Eqs. (1) and (2) will yield pathological results unless 
we derive a basis-space 
dependent $effective$ Hamiltonian: 

\begin{equation}
H^{\Omega}_{eff}=  P \sum_{i=1}^{A}\left [ \frac{\vec{p_i}^2}{2m} + 
\frac{1}{2}m \Omega^2 \vec{r_i}^2 \right ] P 
+ P \left [ \sum_{i<j}^{A}  \left ( V_{ij} - \frac{m \Omega^2}{2A}{(\vec{r_i} - 
\vec{r_j})}^2 \right ) \right ]_{eff} P
\end{equation}

\noindent The second term in Eq. (3) is an effective interaction that is,
in general, an $A$-body interaction. 
If the effective interaction is determined without any approximations, 
the basis-space Hamiltonian  provides an identical 
description of a subset of states as the exact original 
Hamiltonian \cite{PRC48}. From among the
eigenstates  of the Hamiltonian (3), it is necessary to choose 
only those that correspond to the same
center-of-mass energy.  This can be achieved first by working in a 
complete $N \hbar \Omega$ basis space, and
then by projecting the  center-of-mass eigenstates with energies 
greater than $\frac{3}{2}\hbar \Omega$
(representing spurious center-of-mass motion) upwards in the energy spectrum 
\cite{lawson}. In our case, we do this by adding 
$\beta PH^{\Omega}_{c.m.}P$ to and subtracting $\beta \frac{3}{2} \hbar 
\Omega P$ from equation (3) above. One unit of $H^{\Omega}_{c.m.}$ has 
already been incorporated, as mentioned above, and it is also subtracted 
at this stage. 
The resulting shell-model  Hamiltonian takes the form

\begin{eqnarray}
H^{\Omega}_{eff~\beta} &=&  P \sum_{i<j}^{A}\left [ \frac{(\vec{p_i}-
\vec{p_j})^2}{2Am} + \frac{m \Omega^2}{2A} 
(\vec{r_i}-\vec{r_j})^2 \right ] P\\  
&& + P \left [ \sum_{i<j}^{A} \left ( V_{ij} - \frac{m \Omega^2}{2A}{(\vec{r_i} - 
\vec{r_j})}^2 \right ) \right ]_{eff} P
+ \beta P(H^{\Omega}_{c.m.}-\frac{3}{2}\hbar \Omega)P\nonumber
\end{eqnarray}

\noindent where $\beta$ is a sufficiently large positive parameter.

When applied in a suitably chosen finite $HO$ basis space, this procedure 
removes the spurious center-of-mass motion, and has no 
effect on the intrinsic spectrum of states \cite{Petr96}.  Our
choice of the $P$-space is fixed by 
$N \hbar \Omega$ that signifies a many-body basis space of all $HO$ Slater 
determinants having total $HO$ quanta less than or equal to the cutoff given by $N$.
We refer to this as a complete $N \hbar \Omega$ basis space.  It is complete
in the sense it allows the precise removal of spurious center-of-mass motion effects.

In principle, the effective interaction introduced in Eqs. (3) and (4) above 
should reproduce exactly the full-space results in the finite basis space 
for some subset of physical states.  Furthermore, an $A-$body effective 
interaction is required for an $A-$nucleon system. In practice, however, the 
$A-$body effective interaction cannot be calculated exactly, and it is approximated 
by a two-body effective interaction determined for a 
two-nucleon sub-system or two-nucleon cluster. 

In this work, we follow the procedure described in Refs. \cite{Petr96, 
PRLC00}
in order to construct the  two-body effective interaction. The procedure
employs the Lee-Suzuki
\cite{LS1} similarity transformation method, which yields an 
interaction in the form 

\[P_2V_{eff}P_2=P_2VP_2 + P_2VQ_2\omega P_2,\]

\noindent with $\omega$ the transformation operator satisfying 
$\omega=Q_2\omega 
P_2$, and $P_2$, 
$Q_2=1-P_2$ operators that project on the two-nucleon finite basis and complementary 
spaces, respectively. 
Note that we distinguish between the two-nucleon system projection operators 
$P_2,~Q_2$ and the 
$A-$nucleon system projection operators $P,~Q$. Note also that the size of the two-nucleon basis is
dictated by the choice of $N \hbar \Omega$ for the A-body problem.

Our calculations start with exact solutions of the two-body Hamiltonian 

\begin{eqnarray}
H^{\Omega}_2 & \equiv & H^{\Omega}_{02} + V^{\Omega}_2\\
&=& \left [ \frac{\vec{p_1}^2+\vec{p_2}^2}{2m} + \frac{1}{2}m \Omega^2 
(\vec{r_i}^2+ \vec{r_2}^2) \right ] 
+ \left [ V_{12}(\vec{r_1} - \vec{r_2}) - \frac{m \Omega^2}{2A}(\vec{r_1} - 
\vec{r_2})^2 \right ],\nonumber
\end{eqnarray}

\noindent which follows from reducing the shell-model Hamiltonian (2) to the 
two-nucleon case. The exact solutions are obtained by assuming a
0$\hbar \Omega$ motion of the center-of-mass of nucleons 1 and 2. 
Let $|\alpha_P \rangle$ denote the two-nucleon 
$HO$ states which form the basis space signified by $P_2$, and let 
$|\alpha_Q \rangle$ denote those that form the $Q_2-$space. Dropping the subscript ``2'' 
for convenience, one can then express 
the $Q-$space components of an eigenvector $|k \rangle$ of the Hamiltonian (5) as a combination 
of the $P-$space components with the help of the operator $\omega$:

\begin{equation}
\langle \alpha_Q | k \rangle = \sum _{\alpha_P} \langle \alpha_Q | \omega | 
\alpha_P \rangle \langle \alpha_P 
|k \rangle
\end{equation}

\noindent If the dimension of the basis space is $d_P$, we may 
then choose a set 
$\it K$ of $d_P$ eigenvectors 
for which relation (6) will be exactly satisfied \cite{Viaz01}. Typically, 
these $d_P$ states will be the lowest states obtained 
in a given channel. Under the condition that the $d_P \times d_P$ matrix 
$\langle \alpha_P | k \rangle$ for 
$|k \rangle$ $\epsilon$ $\it K$ is invertible, the operator $\omega$ can be 
determined from Eq. (7), and the 
effective Hamiltonian can then be constructed as follows:

\begin{eqnarray}
\langle \gamma_P | H_{2eff} | \alpha_P \rangle =  \sum _{k ~\epsilon~ \it K} 
\left [ \langle \gamma_P |k 
\rangle E_k \langle k| \alpha_P \rangle + \sum _{\alpha_Q}\langle \gamma_P | k 
\rangle E_k \langle k| 
\alpha_Q \rangle \langle \alpha_Q | \omega | \alpha_P \rangle \right ]
\end{eqnarray}

\noindent This Hamiltonian, when diagonalized in a finite basis space, 
reproduces exactly the set $\it K$ of 
$d_P$ eigenvalues $E_k$ of the defined 2-body problem. Note that the 
effective Hamiltonian (7) is not
Hermitian, and that it can be  transformed into a Hermitian form 
$\bar{H}_{2eff}$ by applying a similarity
transformation determined from  the metric operator 
$P_2(1+\omega^{\dagger}\omega)P_2$ \cite{LS2}:

\[\bar{H}_{2eff}=[P_2(1+\omega^{\dagger}\omega)P_2]^{1/2}H_{2eff}
[P_2(1+\omega^{\dagger}\omega)P_2]^{-1/2} \]

The two-body effective interaction actually used in the present 
calculations is determined from this two-nucleon 
effective Hamiltonian as $V_{2eff}=\bar{H}_{2eff}-H^{\Omega}_{02}$. The 
resulting two-body effective interaction 
$V_{2eff}$ depends on $A$, on the $HO$  
frequency $\Omega$, and on $N_{max}$, the maximum many-body 
$HO$ excitation energy (above the lowest 
configuration) defining 
the $P-$space. Furthermore, when 
used in the shell-model Hamiltonian (4), it results in the 
factorization of our 
many-body wave function into 
a product of a center-of-mass $\frac{3}{2}\hbar \Omega$ component times an 
intrinsic component, which allows 
exact correction of any observable for spurious center-of-mass effects, thus 
preserving translational invariance. 
This feature distinguishes our approach from most phenomenological shell-model 
studies that involve multiple $HO$ shells. 

So far, the most important approximation used in our approach is the 
neglect of 
contributions coming from higher 
than two-body clusters to our effective Hamiltonian. Although 
our method can be 
readily generalized to include 
the effects of three-body clusters, leading to the derivation of a three-body 
effective interaction, so far 
this has only been done in $A=3$ and $A=4$ systems 
\cite{Petr98, Petr00}, and it 
is not yet clear how difficult it will be to extend it to the $p-$shell nuclei 
with $A \ge 5$ and with the same range of $N \hbar \Omega$ values accessible
at the 2-body cluster level.

In order to approximately account for the many-body effects 
neglected when using
only a two-body effective  interaction, we introduce a simple
residual interaction 
that depends on all particles.  More specifically, 
our residual many-body 
interaction contains a delta function times the sum of the 
oscillator quanta for the $A-2$ spectators denoted by $N_{sps}$ and defined by 

\begin{equation}
N_{sps} \equiv N_{sum}-N_{\alpha}=N'_{sum}-N_{\gamma}
\end{equation}

\noindent where $N_{sum}$ and $N'_{sum}$ are the total oscillator 
quanta in the initial and final many-body states, respectively. The quantities
$N_{\alpha}$ and $N_{\gamma}$ are the total oscillator quanta in the initial 
and final two-nucleon states 
$| \alpha \rangle$ and $| \gamma \rangle$, respectively, interacting through
the delta function. 

Why do we introduce such a dependence of the many-body forces? We are motivated by the 
experience with the ``multi-$G$'' method of addressing effective many-body forces 
\cite{PRC95}, which is based on the role of the spectators in a many-body matrix element.
As the spectators are excited, the effective $P_2$ space for the interacting pair decreases 
and this increases the attraction of the $G$-matrix. The spectator dependence in the
multi-$G$ approach to the effective many-body forces predominantly lowers the 
excited multi-$\hbar \Omega$ states.

In this connection, one might imagine, as we do, the spectator dependence 
to be of a form where the strength is weakest when the spectators have 0$\hbar \Omega$ 
excitation (above their minimum value) and strongest when they have the maximum
excitation.  These two situations correspond, respectively, to when the
2-particles roam over the entire $N\hbar \Omega$ and over only the 
lowest $\hbar \Omega$ 
range. Thus, we expect this $N_{sps}$-dependent term to be 
attractive and account for the leading effects observed in the multi-$G$ 
investigations.

We also allow for a spectator-independent term in the many-body force and we will
take it to be a delta function for simplicity.  We expect this term will 
approximate the many-body effective force components that are largely 
independent of the spectator configurations.  We expect this contribution to be repulsive in
order to correct for the overbinding found in smaller model spaces when $H_{eff}$ is
approximated with the 2-body cluster form \cite{PRLC00}. We will see below that the fits to 
nuclear spectra verify these expectations.

Our total effective many-body interaction is then expressed in the form of a 
zero-range two-body interaction, 
and two parameters, one of which multiplies the 
many-body variable $N_{sps}$ introduced above:

\begin{equation}
V_{res}(N_{sps}) \equiv (a + b N_{sps})\times G\delta(\vec{r_1}-
\vec{r_2})
\end{equation}

\noindent where $G$ is a fixed strength parameter taken to be 1 $MeV~fm^3$ in the present 
work. The parameters $a$ and $b$ are basis-space dependent dimensionless parameters to be
determined by fits to selected many-body observables. 

We will then use this schematic residual interaction to study the 
role played by the many-body forces that were neglected in our derivation of the effective 
two-body interaction. We shall derive the latter from the realistic Argonne $V8'$ $NN$ 
potential \cite{Av8'}, and perform calculations of some basic observables in the 
$A=4$ system with and without the addition of the residual many-body interaction 
$V_{res}(N_{sps})$, and in various basis spaces. We solve for the properties of $A=4$
nuclei using the $m-$scheme Many-Fermion Dynamics code \cite{MFD} that was developed
for No-Core Shell-Model ($NCSM$) calculations.

\section{Application to $A=4$ Nuclei}

In this section we apply the methods outlined in Sec. II to derive the effective two-body 
interaction, based on the Argonne $V8'$ $NN$ potential, which we then use in the
remainder of our work.  However, it is now well-known that two-body interactions 
alone are not sufficient, and in order to better account for such experimental facts 
as the binding energy of nuclei, one is led to introduce three-body and higher multi-body 
forces. In the following discussion, we first consider calculations done 
with only the Argonne $V8'$ two-body interaction in order to see how such two-body 
calculations compare with experiment. We then ask the following question: what couplings 
$(a,b)$ for 
the residual many-body interaction are needed in order to achieve as close an agreement with 
experiment as possible? 

We use a complete $N \hbar \Omega$ basis space with, e.g., $N=8$ for the 
positive-parity states of $^4He$. This means that nine major $HO$ shells are employed. 
The two-nucleon basis space is defined in our calculations by $N_{max}$, e.g., for an 
8$\hbar \Omega$ calculation for 
$^4He$ we have $N_{max}$=8. The restriction of the $HO$ shell 
occupation is given by $(N_1 + N_2) \leq N_{max}$ where the single-particle
$HO$ quanta are given by $N_i = 2 n_i + l_i$.
The  same conditions hold for the 
relative $2n+l$, center-of-mass $2 N + L$, and $2n+l+2 N +  L=N_1+N_2$ 
quantum numbers, respectively. 

In the case of the $A=4$ system, we seek to reproduce the following obervables 
as accurately as possible: the binding energy of the ground state ($GS$) as well 
as the energies of the first three excited states. We are also interested in reproducing 
the root-mean-square radius $R$.

\subsection{The Experimental Situation}

In Table 1 we present the experimental energies of the low-lying states that are relevant 
to our present study. The binding energies of the ground states are given first, 
followed by the excitation energies of the first three excited states (all energies are 
in $MeV$)  \cite{he4exp, ndata}. Furthermore, the root-mean-square point-charge radius 
of the ground state of $^4He$ is known experimentally to be 1.46 $fm$, which we also take 
to be the point mass radius. 

\subsubsection{Calculations Using Only Two-Body Effective Interactions}

Before we embark upon the investigation of the many-body force, 
let us first examine the theoretical predictions of the $^4He$ spectrum using 
only the Argonne $V8'$ plus the Coulomb interaction. In Table 2, we present
results of calculations  of the first two positive-parity and the first two 
negative-parity states, using the  interaction in increasingly larger basis  
spaces (indicated by $N_{max}$, as defined above). The results 
depend somewhat 
on the choice of $\hbar \Omega$, the $HO$ energy parameter,
but we shall adopt the value $\hbar \Omega$=19 $MeV$, which is suggested 
by the phenomenological expression \cite {Ber72}

\[\hbar \Omega = \frac{45}{A^{1/3}}~-~\frac{25}{A^{2/3}}. \]

\noindent In the large $N_{max}$ limit our results become  
independent of the choice of $\hbar \Omega$ \cite {Bar01}.

As shown in Table 2, it is clear that the Argonne $V8'$ 
interaction overbinds the ground 
state for $N_{max} < 8$ and underbinds it for 
$N_{max}=8$ and 10, with the overbinding decreasing 
(and the underbinding increasing) as $N_{max}$ 
increases. On the other hand, the excitation energy 
of the $0^+_2$ state steadily $decreases$ with $N_{max}$. If 
a two-body force is alone sufficient, 
and if the Argonne $V8'$  plus Coulomb interaction 
represents a sufficiently accurate form of this
two-body force, we would expect our calculations to 
converge to the experimental 
values. By examining Table 2, it  becomes clear that 
while the first positive-parity 
excited state ($0^+_2$) and the negative-parity  
states ($0^-$ and $2^-$) are becoming more 
bound with increasing $N_{max}$, the  ground state 
itself ($GS$) is becoming less bound. 

Furthermore, while the discrepancy between the calculated and the experimental ground-state 
energy is reduced with increasing $N_{max}$ (from 12\% with $N_{max}=2$ to 
less than  4\% with $N_{max}=10$), the improvement is less favorable for the energies of the 
excited states ($e.g.,$ from 51\% with $N_{max}=2$ to 23\% with $N_{max}=10$ for the $0^+_2$ 
state). The percentages used here represent the percent error (in absolute value) of the 
calculated energy relative to the experimental value. For the ground state, this is based 
on the total binding energy, whereas for the excited states it is based on the excitation 
energy.
  
We are, thus, led to conclude that the two-body force used is somewhat deficient 
and we need either a better two-body force, or three-body and/or higher 
many-body forces, in general. This is exactly the situation that we wish to investigate with 
our zero-range, spectator-dependent, residual delta interaction. 

\subsubsection{The effect of adding a residual many-body force}

Next, we consider the effect of adding our residual delta interaction to the derived 
effective 2-body interaction in order to obtain a 2 + many-body potential: 

\[V=V^{eff}_{Argonne V8'} + V_{Coul} + [a + b N_{sps}]\times
G\delta(\vec{r_1}-\vec{r_2}). \]	

\noindent By adding the spectator-dependent phenomenological interaction introduced in 
Eq. (9), we are simply modelling the contribution of all residual many-body forces, omitted 
by our derived two-body effective interaction. We then vary the coefficients $a$ and $b$ 
independently for each basis space, in order to achieve good agreement with the
experimental ground-state binding energy and excitation energy of the first
excited $0^+$ state  of $^4He$. The results are shown in Table 3 and illustrated in
Fig. 1. We note that $a$ is  positive, while $b$ is negative, and that both are
decreasing functions of $N_{max}$, with 
$a$ decreasing somewhat faster than $b$. As the ground state was found to be 
overbound, whereas the first excited state was found to be 
underbound by the pure two-body interaction (see Table 2), it should not be surprising 
that $a$ turns out to be positive, while $b$ is negative. It is also reassuring to see 
both $a$ and $b$ decrease steadily (in absolute value) with $N_{max}$, indicating that 
the need for the residual many-body interaction $V_{res}$ decreases and that the 
adequacy of the two-body interaction increases as the basis space becomes larger. More 
precisely, the result that $a \rightarrow 0$ as $N_{max} \rightarrow \infty$, is not 
surprising, as it indicates that there is little or no correction needed at the pure two-body 
level to our Hamiltonian. On the other hand, the $b$-term, which carries all the 
information about the "true" many-body interaction, is $not$ converging to zero. We, therefore, 
conclude that it is possible to accommodate the contribution of the missing many-body 
interaction and to represent it in the simple form of $b N_{sps}\times 
G\delta(\vec{r_1}-\vec{r_2})$ for these observables.

We are now in a position to examine the predictive power of our extended effective 
Hamiltonian. We begin by asking how well such a residual many-body interaction (determined 
in the case of $^4He$) can help reconcile the calculated spectra of $^4H$ and $^4Li$ 
with the experimental situation? In Tables 4 and 5 we show the results obtained with and 
without the residual many-body interaction and compare them to experiment. We are 
now addressing ($T=1$) negative-parity states, so we perform our calculations 
in an $N_{max}=9$ basis space, and we use the values $a=0.63$ and $b=-2.43$, which were 
determined in the earlier fit to $^4He$ (see Table 3). 

It is clear from the results shown in Tables 4 and 5 that a much better 
agreement with experiment is achieved for both $^4H$ and $^4Li$ when our 
residual many-body interaction is added to the pure two-body interaction. 
Aside from the improved agreement for the ground states and first excited 
states, we observe that it is only by including the residual many-body interaction that the 
calculated $0^-$ and $1^-_2$ states in $^4Li$ become bound, in accord with experiment. 

Based on the overall results for the $A = 4$ isotopes, presented above, we 
can identify a procedure for improving the many-body force. In particular, the $^4He$ fit can 
be interpreted as determining the spin-isospin averaged values of the couplings for our 
many-body force. By allowing spin-isospin dependence in the  couplings, we could fine tune the
coupling strengths, while preserving their  average values, so as to obtain an improved
description of the other isotopes.  Thus, we can introduce an isospin dependence in our residual
many-body force as follows:

\[V_{res} = [(a+a'\vec{\tau_1}\cdot \vec{\tau_2}) + (b+b'\vec{\tau_1}\cdot \vec{\tau_2}) 
N_{sps}]\times G\delta(\vec{r_1}-\vec{r_2}). \]	

\noindent or, equivalently, by assigning different values for $a$ and $b$ in the $T=0$ and $T=1$ 
channels, denoted by $a_0,~a_1$ and $b_0,~b_1$. We do this while keeping the isospin average 
of $a_0$ and $a_1$ to be equal to $a$, and likewise for $b_0$ and $b_1$. Applying a minimum 
root-mean-square convergence criterion on the calculated energies of the first two states 
in each of the $A=4$ nuclei considered relative to their experimental values, we find that it 
is possible to improve the calculations when using an isospin-dependent version of the residual 
many-body interaction (Table 6).

It would be interesting to study whether our model of the residual many-body interaction can be 
generalized to the study of $A > 4$ systems. For example, the test of the spin-isospin 
dependent many-body force can then be carried out in the A = 5 systems. This effort is underway 
and will be reported in another paper. 

To summarize, we have presented a simple model for treating multi-body interactions in 
nuclear-structure calculations in terms of a two-parameter, basis-space-dependent delta 
interaction. We determine the two-parameters as a function of basis-space size by fits to 
the lowest two $0^{+}$ states of $^4He$. The parameter values decrease with basis-space size, 
as anticipated. The obtained multi-body interaction also yields improved agreement with 
experiment for negative parity states in $^4He$ and for the T = 1 states of $^4H$ and 
$^4Li$, which were not included in the empirical fits. The advantages of our model for the 
multi-body interaction in nuclei are two-fold: (1) the model provides a simple method for 
including multi-body interactions into nuclear-structure calculations, thereby yielding 
improved agreement with experiment, and (2) the model will give us insight into the nature of 
such multi-body interactions, so that we will be able to model them more realistically
in the future.

\section{Acknowledgements}

M.S. Fayache and B.R. Barrett acknowledge 
partial support from NSF Grant No. PHY0070858.
J.P. Vary acknowledges support from USDOE Grant No. DE-FG02-87ER-40371.  
This work was partly performed under the auspices of
the U. S. Department of Energy by the University of California,
Lawrence Livermore National Laboratory under contract
No. W-7405-Eng-48 (P. Navr\'{a}til).

\pagebreak

\pagebreak

{\large{\bf {Figure Captions\\}}}

{\bf {Fig. 1:}} The strengths $a$ (solid curve) and $b$ (dashed curve) 
of the residual delta interaction needed to achieve 
agreement with experiment in the 
case of $^4He$ and using $\hbar \Omega=19~MeV$.\\

\pagebreak

\begin{table}
\caption{Experimental binding energies of the ground states ($GS$) 
and excitation energies of the first three excited states (in $MeV$) in the $A=4$ system. }
\begin{tabular}{cc|cc|cc}
\hline                                                   
  \multicolumn{2}{c|}{$^4He$} & 
  \multicolumn{2}{c|}{$^4Li$}&
  \multicolumn{2}{c}{$^4H$}\\
  \multicolumn{2}{c|}{$(T=0)$} & 
  \multicolumn{2}{c|}{$(T=1)$}&
  \multicolumn{2}{c}{$(T=1)$}\\
\hline
$0^+_1$ ($GS$) & -28.295674 5 & $2^-_1$ ($GS$) & -4.618058 & $2^-_1$ ($GS$) & -5.575154\\
        &        &         &       &         & \\
$0^+_2$ & 20.210 & $1^-_1$ & 0.320 & $1^-_1$ & 0.310 \\
        &        &         &       &         & \\
$0^-_1$ & 21.010 & $0^-_1$ & 2.080 & $0^-_1$ & 2.080\\
        &        &         &       &         & \\
$2^-_1$ & 21.840 & $1^-_2$ & 2.850 & $1^-_2$ & 2.830\\
\hline
\end{tabular}  
\end{table}
 
\begin{table}  
\caption{Ground-State energy and excitation energies (in $MeV$) 
of the first two $0^+$ states and the first two negative-parity states 
of $^4He$ using $only$ the Argonne $V8'$ and the Coulomb potentials ($i.e.,~a=0$ and $b=0$), 
as calculated in shell-model spaces characterized  by $N_{max}\hbar \Omega$ 
excitations and with $\hbar \Omega=19~MeV$.}

\begin{tabular}{ccc|ccc}
\hline                                                   
  \multicolumn{3}{c|}{(a) Positive Parity} & 
  \multicolumn{3}{c}{(b) Negative Parity}\\
\hline
 $N_{max}$ & $E(GS)$ & $E_x(0^+_2)$ &  $N_{max}$ & $E_x(0^-)$ & $E_x(2^-)$ \\
\hline                                                   
 2 & -31.853 & 30.577 & 3 & 26.560 & 28.692\\
   &             &           &   &   &  \\
 4 & -30.758 & 29.825 & 5 & 25.694 & 27.600\\
   &             &           &   &   &  \\
 6 & -29.328 & 27.019 & 7 & 23.884 & 25.644 \\
   &             &           &   &   &  \\
 8 & -28.149 & 25.213 & 9 & 22.607 & 24.211 \\
   &             &           &   &   &  \\
10 & -27.247 & 24.902 &  &  & \\
\hline
\end{tabular}  
\end{table}

\pagebreak

\begin{table}  
\caption{Values of the parameters $a$ and $b$ in the interaction 
$V^{eff}_{Argonne V8'} + V_{Coul} + [a~+~b *N_{sps}]\times G\delta(\vec{r_1}-\vec{r_2})$ and 
corresponding energies as calculated in various basis spaces characterized by $N_{max}\hbar 
\Omega$ excitations, with $\hbar \Omega=19~MeV$ for $^4He$.}

\begin{tabular}{cc|cccc|ccc}
\hline                                                   
$a$ & $b$ & \multicolumn{4}{c|}{(a) Positive Parity} & 
  \multicolumn{3}{c}{(b) Negative Parity}\\
\hline
 & &  $N_{max}$ & $E(GS)$ & $R_{GS}$ & $E_x(0^+_2)$ &  $N_{max}$ & $E_x(0^-)$ & 
$E_x(2^-)$ \\
\hline                                                   
 2.22  &  -6.86 & 2 & -28.305  & 1.390 & 20.223  & 3 & 23.811 & 
26.017\\
   &   &  &           &  &         &   &   &  \\
2.09 &  -5.80 & 4 & -28.291 & 1.341 & 20.221 & 5 & 23.638 & 
25.353\\
   &             &       &    &   &   &  \\
1.15 &  -3.12 & 6 & -28.296  & 1.354  & 20.212 & 7 & 22.446  & 
23.977 \\
   &             &        &   &   &   &  \\
0.63 &  -2.43 & 8 & -28.301  & 1.383 & 20.238 & 9 & 21.541 & 
22.903 \\
   &             &      &     &   &   &  \\
0.09 &  -1.78 & 10 & -28.295 & 1.409 & 20.209 &  &  & \\
\hline
\end{tabular}  
\end{table}

\begin{table}  
\caption{Experimental and calculated energies (in $MeV$) 
of the first few states 
in $^4H$, with and without the residual many-body 
interaction ($a=0.63,~b=-2.43$) 
in an $N_{max}=9$ basis space and using $\hbar \Omega=19~MeV$.}

\begin{tabular}{cc|c|c}
\hline                                                   
State & Experimental & Energy calculated without &  Energy calculated with\\
      & Energy       & residual many-body interaction & residual many-body 
interaction\\ 
\hline                                                   
$2^-_1~(GS)$ & -5.575 & -2.994 & -4.922 \\
             &              &                    & \\
$1^-_1$      & -5.265 & -2.452  & -4.792  \\
             &              &                    & \\
$0^-_1$      & -3.495 & -1.403  & -3.861  \\
             &              &                    & \\
$1^-_2$      & -2.745 & -0.935  & -3.864 \\ 
\hline
\end{tabular}  
\end{table}

\pagebreak

\begin{table}  
\caption{Experimental and calculated energies (in $MeV$) of the first few states 
in $^4Li$, with and without the residual many-body interaction ($a=0.63,~b=-2.43$) 
in an $N_{max}=9$ basis space, and using $\hbar \Omega=19~MeV$.}

\begin{tabular}{cc|c|c}
\hline                                                   
State & Experimental & Energy calculated without &  Energy calculated with\\
      & Energy       & residual many-body interaction & residual many-body 
interaction\\ 
\hline                                                   
$2^-_1~(GS)$ & -4.618 & -1.409 & -3.433  \\
             &              &                    & \\
$1^-_1$      & -4.298 & -0.933  & -3.343  \\
             &              &                    & \\
$0^-_1$      & -2.538 &  0.097  &  -2.422 \\
             &              &                    & \\
$1^-_2$      & -1.768 &  0.532 &  -2.444 \\ 
\hline
\end{tabular}  
\end{table}

\begin{table}  
\caption{Experimental and calculated energies (in $MeV$) of the first two states 
in the $A=4$ system, with the isospin-independent and the isospin-dependent 
residual many-body
interaction in $N_{max}=8, 9$ basis spaces, and using $\hbar \Omega=19~MeV$.}

\begin{tabular}{cc|c|c}
\hline                                                   
Nucleus (State) & Experimental & Using $a_0=a_1=0.63$ &  Using $a_0=0.52,~ a_1=0.67$\\
      & Energy       & and $b_0=b_1=-2.43$ & and $b_0=-2.88, ~b_1=-2.28$\\
\hline
 
$^4He$($0^+_1~(GS))$ & -28.297 & -28.301 & -28.627 \\
             &              &                    & \\
$^4He$($0^+_2)$      & -8.086 & -8.063 & -8.697  \\
             &              &                   & \\
\hline                                                   
 
$^4H$($2^-_1~(GS))$ & -5.575 &  -4.922 & -5.255\\
             &              &                    & \\
$^4H$($1^-_1)$      & -5.265 &  -4.792 & -5.157  \\
             &              &                    & \\
\hline                                                   

$^4Li$($2^-_1~(GS))$ & -4.618 & -3.433 & -3.775    \\
             &              &                    & \\
$^4Li$($1^-_1)$      & -4.298 &  -3.343 & -3.713 \\
             &              &                    & \\
\hline
\end{tabular}  
\end{table}

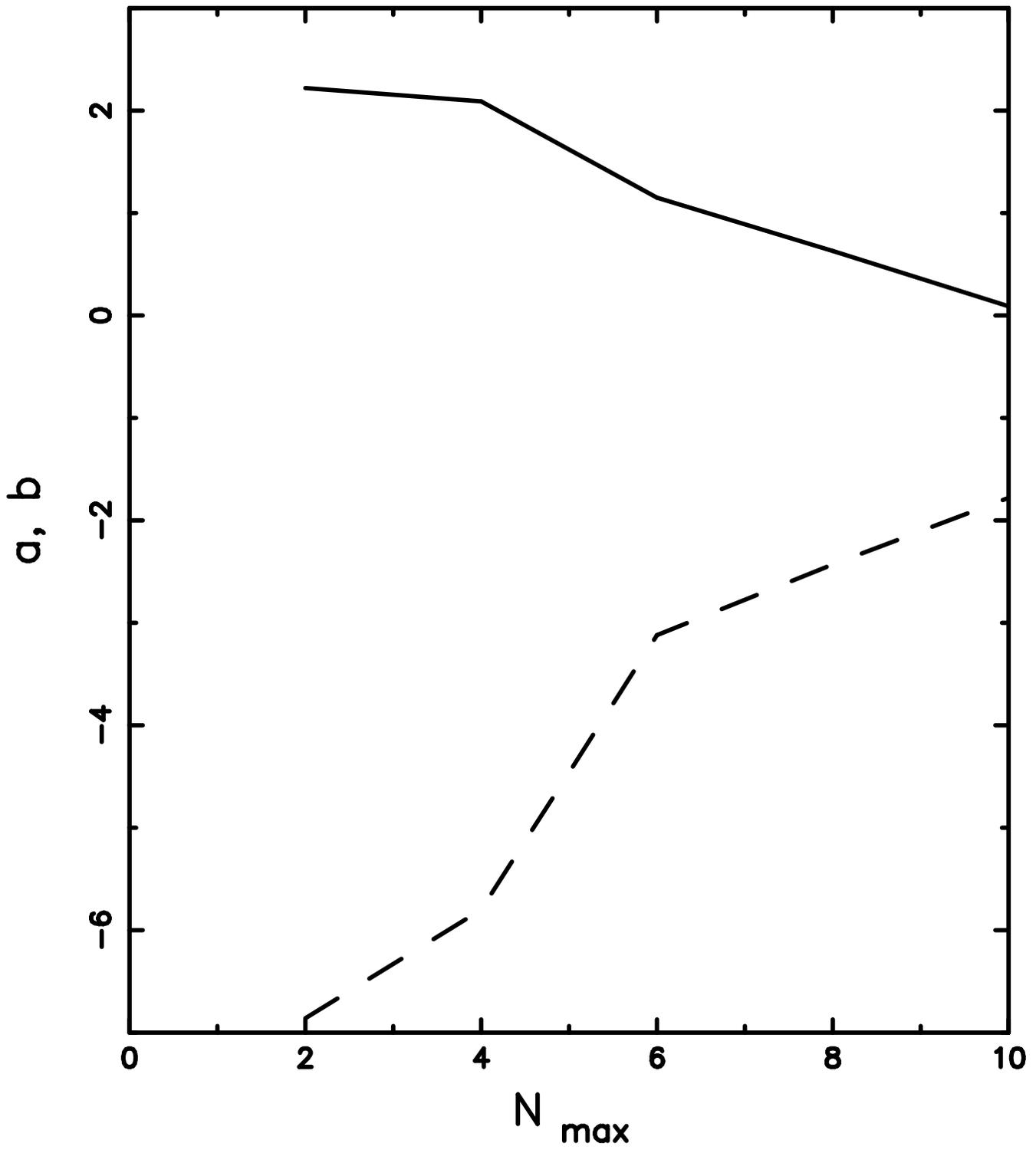
\end{document}